\documentclass[]{spie}  %>>> use for US letter paper
\usepackage{amsmath,amsfonts,amssymb}
\usepackage{graphicx}
\usepackage[colorlinks=true, allcolors=blue]{hyperref}

\usepackage{makecell}
\usepackage{multirow}
\usepackage{makecell}
\usepackage{adjustbox}
\usepackage{float}
\usepackage{graphicx}
\usepackage{subcaption}

\title{Weakly Supervised Detection of \\ Pheochromocytomas and Paragangliomas in CT}

\author[a]{David C. Oluigbo}
\author[a*]{Bikash Santra}
\author[a*]{Tejas Sudharshan Mathai}
\author[a]{Pritam Mukherjee}
\author[a]{\\Jianfei Liu}
\author[b]{Abhishek Jha}
\author[b]{Mayank Patel}
\author[b]{Karel Pacak}
\author[a]{Ronald M. Summers}
\affil[a]{Imaging Biomarkers and Computer-Aided Diagnosis Laboratory, Radiology and Imaging Sciences, Clinical Center, National Institutes of Health, Bethesda MD, USA}
\affil[b]{National Institute of Child Health and Human Development, National Institutes of Health, Bethesda, MD, USA}
\affil[*]{Equal contribution}

\authorinfo{Send correspondence to T.S.M: tejas dot mathai at nih dot gov}

\graphicspath{ 
{Figures/fig_money/} 
{Figures/fig_results/}
}

\begin{document} 
\maketitle

\begin{abstract}
Pheochromocytomas and Paragangliomas (PPGLs) are rare adrenal and extra-adrenal tumors which have the potential to metastasize. For the management of patients with PPGLs, CT is the preferred modality of choice for precise localization and estimation of their progression. However, due to the myriad variations in size, morphology, and appearance of the tumors in different anatomical regions, radiologists are posed with the challenge of accurate detection of PPGLs. Since clinicians also need to routinely measure their size and track their changes over time across patient visits, manual demarcation of PPGLs is quite a time-consuming and cumbersome process. To ameliorate the manual effort spent for this task, we propose an automated method to detect PPGLs in CT studies via a proxy segmentation task. As only weak annotations for PPGLs in the form of prospectively marked 2D bounding boxes on an axial slice were available, we extended these 2D boxes into weak 3D annotations and trained a 3D full-resolution nnUNet model to directly segment PPGLs. We evaluated our approach on a dataset consisting of chest-abdomen-pelvis CTs of 255 patients with confirmed PPGLs. We obtained a precision of 70\% and sensitivity of 64.1\% with our proposed approach when tested on 53 CT studies. Our findings highlight the promising nature of detecting PPGLs via segmentation, and furthers the state-of-the-art in this exciting yet challenging area of rare cancer management.

\end{abstract}

% This model offers an automated solution that can significantly enhance the efficiency of PPGL detection.


\section{INTRODUCTION}
\label{sec:intro}  % \label{} allows reference to this section

Pheochromocytomas and Paragangliomas (PPGLs) are rare neuroendocrine tumors that can appear sporadically in the body or due to various inherited pathogenic variants. Pheochromocytomas originating from chromaffin cells in the adrenal medulla make up 80-85\% of PPGLs \cite{Lenders_2005, Eisenhofer_2003}. Conversely, Paragangliomas can originate from extra-adrenal chromaffin cells located in the abdomen, pelvis, head and neck, or chest regions. The only account for 15-20\% of PPGLs \cite{Lenders_2005, Eisenhofer_2003}. Diagnosing PPGLs is especially important for those that exhibit a higher metastasis rate. For example, trunk paragangliomas can exhibit metastasis rates as high as 60\%, making their detection very relevant for patient outcomes \cite{Wen_2010, Ayala_Ramirez_2011}. Paragangliomas are mostly found in the neck, abdomen and pelvis, and $\sim$2\% originate in the chest \cite{Brown_2008}. CT imaging is the preferred modality for clinicians to localize the PPGLs, track their progression, and determine their metastatic potential. Therefore, it is essential to precisely localize them and monitor their progression as undiagnosed PPGLs can have fatal implications for patients.

However, some intrinsic properties of PPGLs render their detection challenging. For example, these tumors have varying sizes that range from 1 to 15 cm in maximum diameter. They can also vary in their morphology with smaller PPGLs being typically homogeneous, whereas larger PPGLs exhibit central necrosis. Additionally, they can exhibit macroscopic fat, hemorrhage, or calcification that make them more difficult to identify and distinguish from other masses, such as adenomas \cite{Leung_2013,Garcia_Carbonero_2021}. Despite these hurdles, it is clinically relevant to detect PPGLs as they have the potential to improve patient outcomes. Moreover, it can be utilized for downstream tasks, such as the identification of their genetic makeup \cite{nolting2022personalized, noortman202218f}.

%%%%%%%%%%%
\begin{figure}[!htb]
\centering
\includegraphics[width=0.8\textwidth]{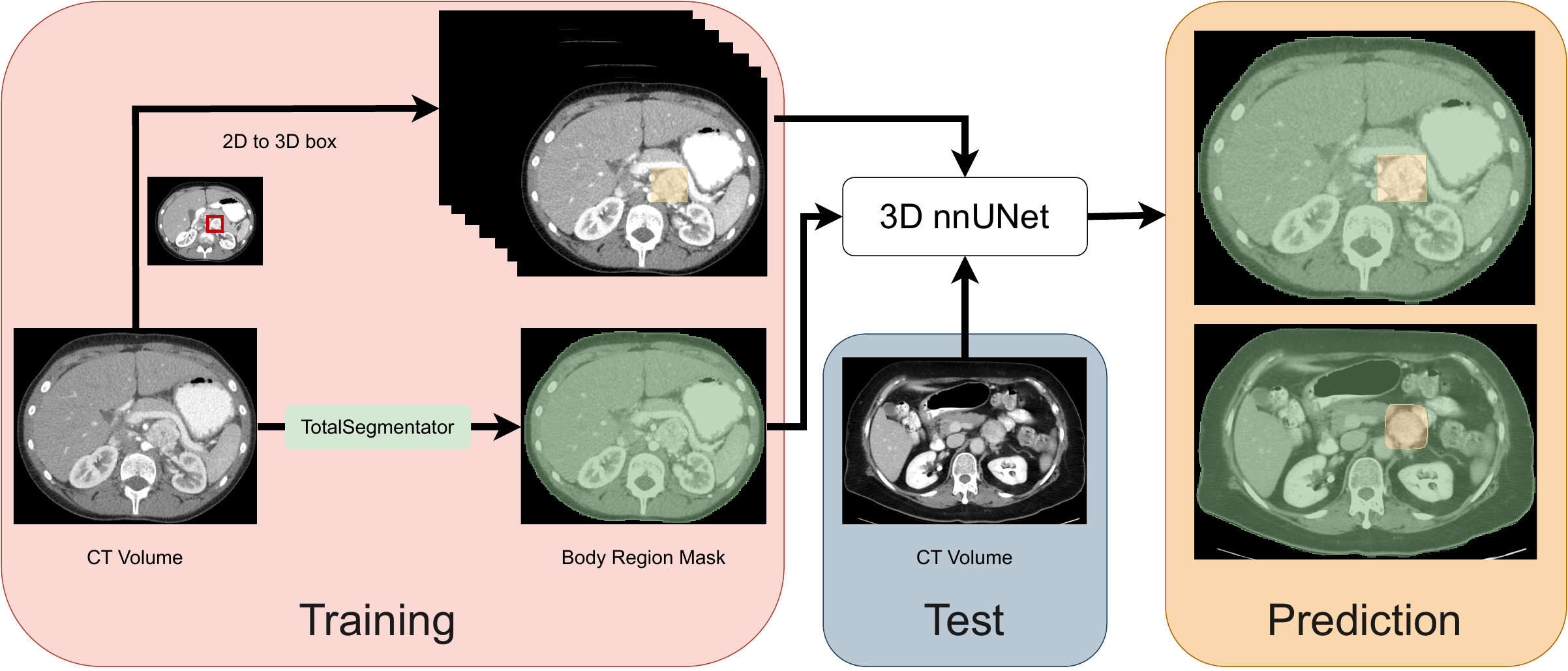}
\medskip
\caption{Framework for the detection of pheochromocytomas and paragangliomas (PPGLs) via a proxy segmentation task using a 3D nnUNet. PPGLs in CT volumes were annotated with 2D boxes (red box), and these were converted into weak 3D segmentations (yellow box). The body region mask (green) from TotalSegmentator was also merged with the weak 3D annotations. The 3D nnUNet model was trained to segment PPGLs annotated in the CT volumes. At test time, the model received a 3D CT volume and detected PPGLs (via segmentation).}
\label{bd}
\end{figure}
%%%%%%%%%%%

In this pilot work, we endeavor to detect PPGLs via a proxy segmentation task. We used the nnUNet \cite{Isensee_2020} framework to segment PPGLs owing to its superior segmentation performance. As the annotation of a tumor in each slice of the CT volume is cumbersome and time-consuming, we used weak 3D bounding box annotations for training nnUNet as illustrated in Fig. \ref{bd}. Our results indicated that the weak annotations were sufficient to train the nnUNet model to detect the PPGLs via segmentation. To the best of our knowledge, this work is the first known attempt to automatically detect and segment PPGLs in CT volumes.

\section{Methods}

\subsection{Dataset}
\label{sec:ds}
The Picture Archiving and Communication System (PACS) at the NIH Clinical Center was queried for patients who underwent CT imaging between 1999 and 2022. Initially, 300 portal venous phase CT scans were collected for 289 patients. Certain patients underwent separate CT scans for their head-neck and chest-abdomen-pelvis regions, respectively. The acquisition of contrast-enhanced CT scans followed a consistent protocol involving a fixed delay of 70-seconds post-intravenous contrast material administration. The scans had varying voxel spacing ranging from 1 to 10 mm. A total of 1010 PPGLs were found in the 300 CT scans. As the focus of our work was to detect paragangliomas in the body, scans containing tumors in the head or neck (n=42) were excluded, yielding 258 scans (one scan per patient). We also excluded scans where no lesions had been observed (n=3). The remaining 255 CT scans were used in this work, and they were divided into training ($\sim$80\%, 202 scans) and test ($\sim$20\%, 53 scans) splits. The 53 test CT scans contained 153 PPGLs.

% (63.5\%, 162 scans), validation (15.7\%, 40 scans), and test (20.8\%, 53 scans) splits.

\subsection{Ground-Truth Generation}

\textbf{PPGL annotation.} In clinical practice, radiologists routinely scroll through the CT volume to identify abnormal findings. They measure the tumor extent (using RECIST measurements) in only one slice as the manual measurement in all slices is time-consuming and cumbersome. Furthermore, if many tumors are found, they only annotate a few ``significant'' tumors depending on their size ($\geq$ 1cm). In our pilot work, we replicated this process of annotation for the PPGLs. First, the CT volume was loaded into ITK-SNAP \cite{py06nimg} with a window center and width of [50, 450] HU. A trained grader visually determined the axial slice showing the maximum diameter of a PPGL, and a bounding box was drawn around each tumor on the chosen slice. This served to mimick the prospective annotations performed by the radiologists. These 2D boxes were then verified by a senior board-certified radiologist (R.M.S) with 30+ years of experience. Then, we extended the 2D box to the adjacent three slices above and below the current slice. The regions covered by the 3D boxes were designated as containing PPGLs. This was a cost-effective and time-efficient manner to create the weak 3D boxes for training the model.

\noindent
\textbf{Body Region Segmentation.} Since PPGLs can either be small or large, a nnUNet model trained directly with the 3D boxes as segmentation masks will have degraded performance due to a class imbalance between the foreground (PPGL) and background (all other regions) class. To combat the class imbalance issue, we also utilized the segmentation of the body region as shown in Fig. \ref{bd}, which was generated using the TotalSegmentator \cite{Wasserthal_2023} tool. This tool produced segmentation masks for various organs in CT, and it distinguished the body region from the background in the CT. The PPGL 3D box segmentation masks were merged with the body region segmentations.

\subsection{Detection of PPGL using nnUNet} 

The nnUNet \cite{Isensee_2020} is a self-configuring segmentation framework that can be adapted to different datasets and modalities, such as CT. It automatically determined the optimal hyper-parameters for training a segmentation model and learned to segment target structures of interest. We trained a 3D full-resolution nnUNet for detecting PPGLs via a proxy segmentation task. During training, our pipeline took a CT volume and its corresponding ground-truth mask as input. The nnUNet model learned to generate a segmentation for the CT volume and iteratively refined it via a loss function that computed a segmentation error that measured the overlap between the prediction and ground-truth. At inference time, nnUNet predicted the segmentation mask for an input CT volume, and they included the PPGLs present in the volume along with the body region mask. 

% \begin{figure}[!t]
% \centering
% \includegraphics[width=0.9\textwidth]{SPIE_Figure1_Revised.pdf}
% \caption{Rows 1, 2, and 3 show example axial CT slices, ground-truth PPGLs (yellow boxes), and detected PPGLs (blue boxes) overlaid, respectively. Columns 1, 2, and 3 show true positives; a small part of an oblong tumor in column 3 was detected by nnUNet. In Column 4, a false positive incorrectly predicted by nnUNet is shown. 
% % Column 4 shows a tumor that was correctly detected, but the GT was incomplete and it rendered this prediction to incorrectly be considered as a false positive.
% }
% \end{figure}

\begin{figure}
\centering
\begin{subfigure}{0.235\textwidth}
    \centering
    \includegraphics[width=\linewidth,height=2.6cm]{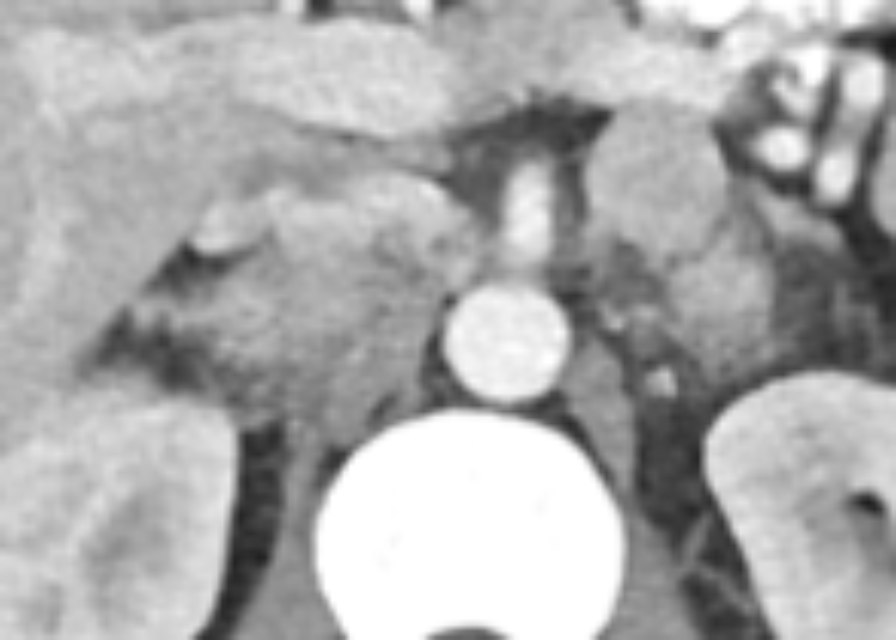}
    \smallskip
    \includegraphics[width=\linewidth,height=2.6cm]{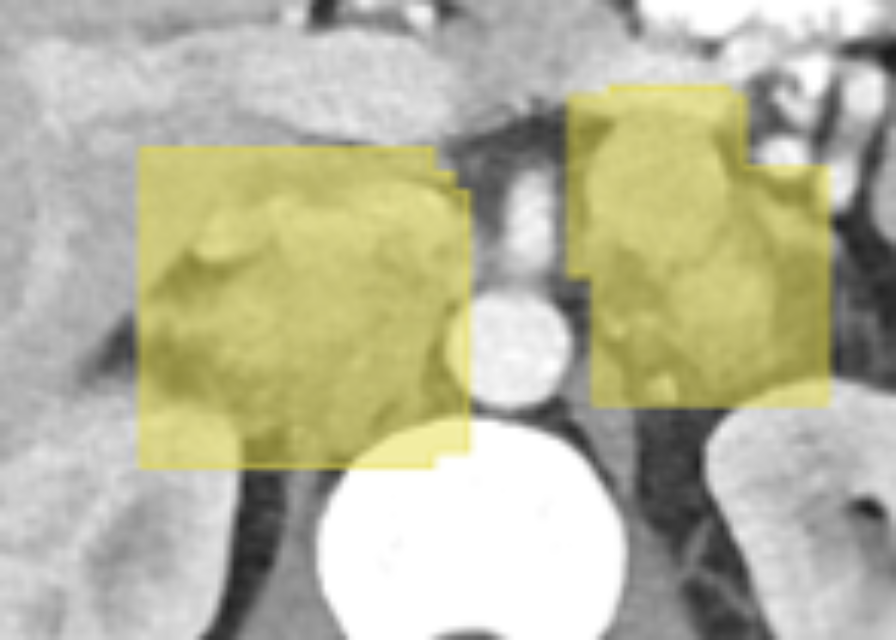}
    \smallskip
    \includegraphics[width=\linewidth,height=2.6cm]{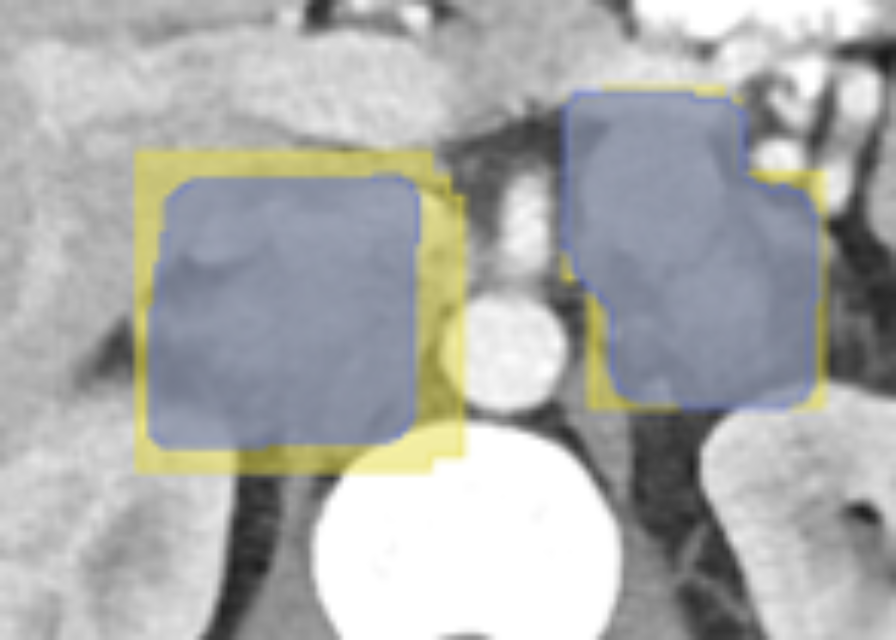}    
    %\caption{}
\end{subfigure}
\begin{subfigure}{.235\textwidth}
    \centering
    \includegraphics[width=\linewidth,height=2.6cm]{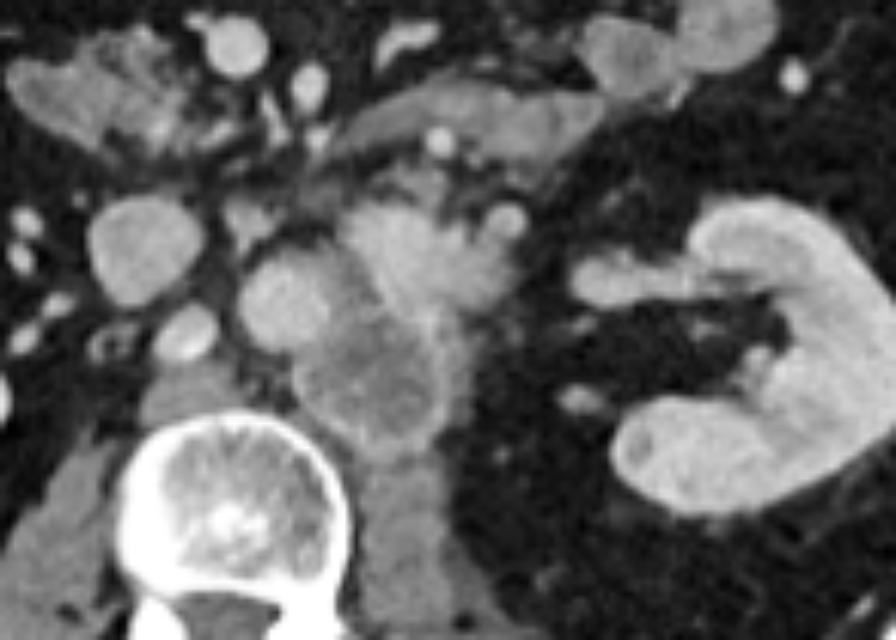}
    \smallskip
    \includegraphics[width=\linewidth,height=2.6cm]{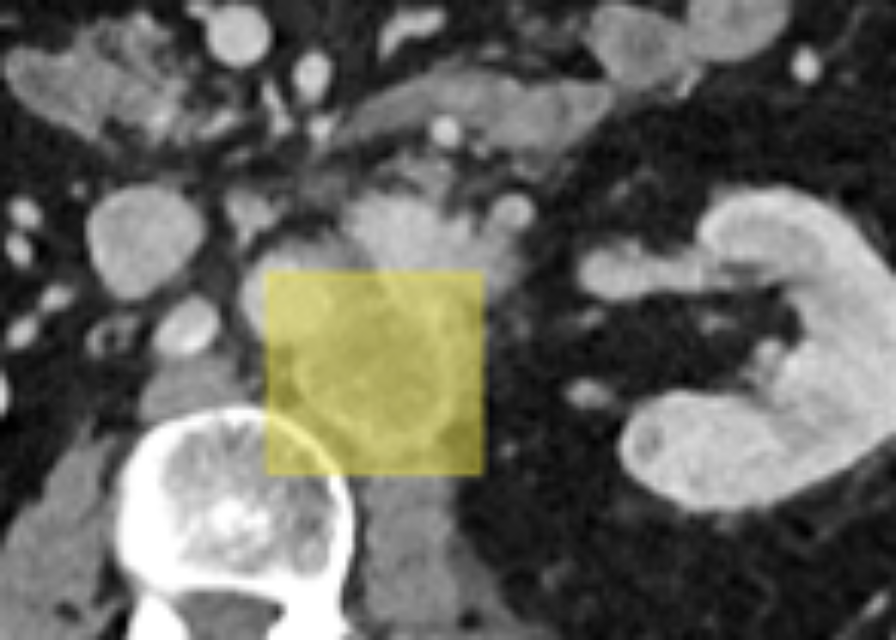}
    \smallskip
    \includegraphics[width=\linewidth,height=2.6cm]{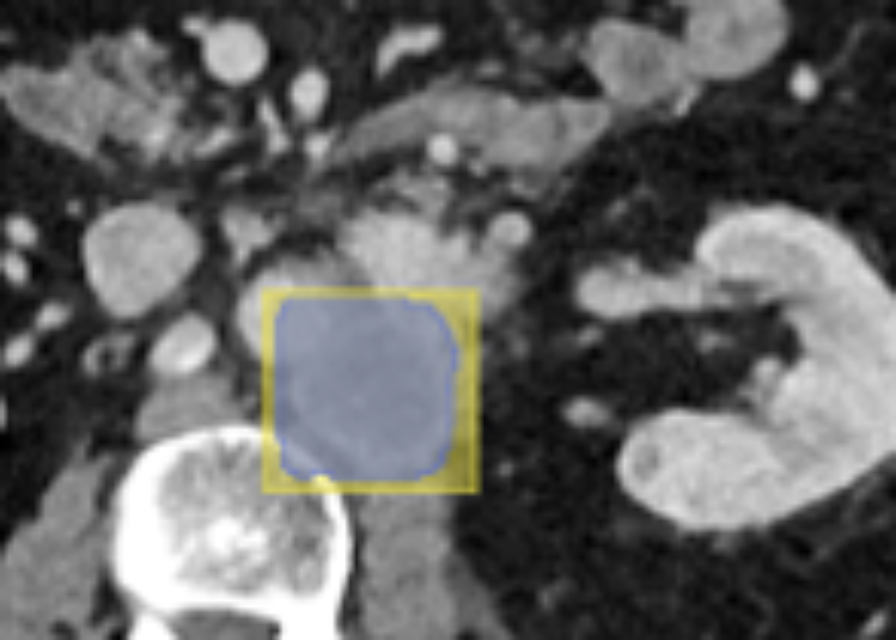}    
    %\caption{}
    
\end{subfigure}
\begin{subfigure}{.235\textwidth}
    \centering
    \includegraphics[width=\linewidth,height=2.6cm]{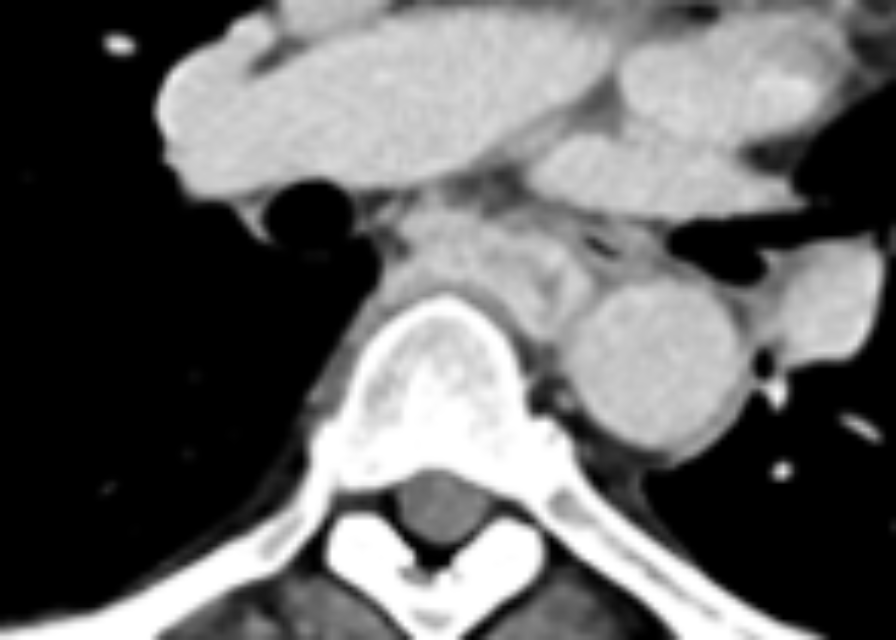}
    \smallskip
    \includegraphics[width=\linewidth,height=2.6cm]{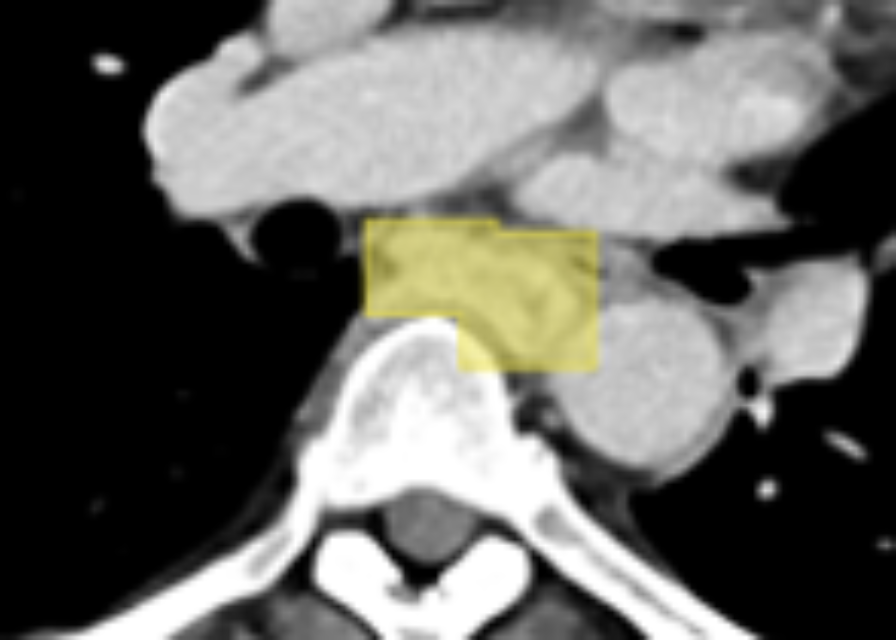}
    \smallskip
    \includegraphics[width=\linewidth,height=2.6cm]{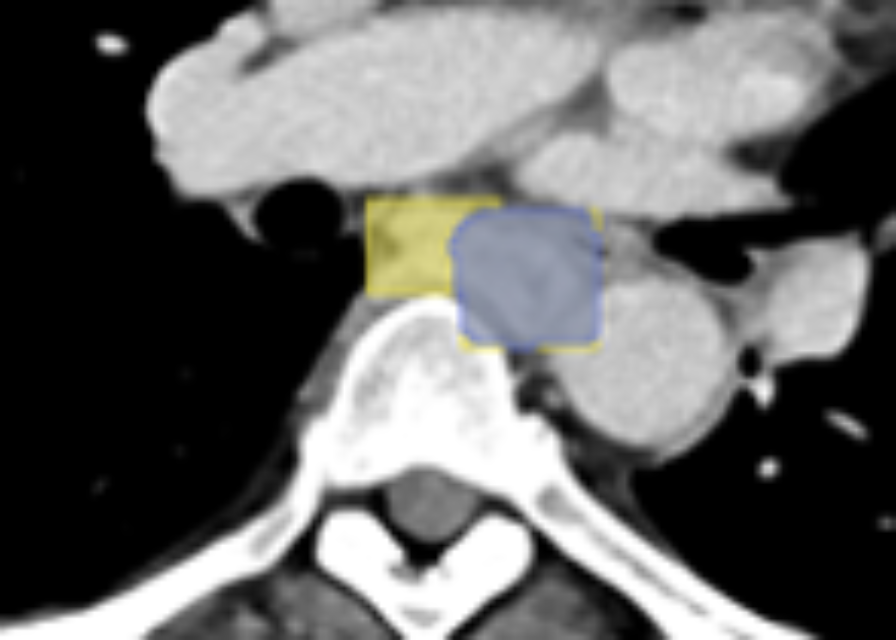}    
    %\caption{}
    
\end{subfigure}
\begin{subfigure}{.235\textwidth}
    \centering
    \includegraphics[width=\linewidth,height=2.6cm]{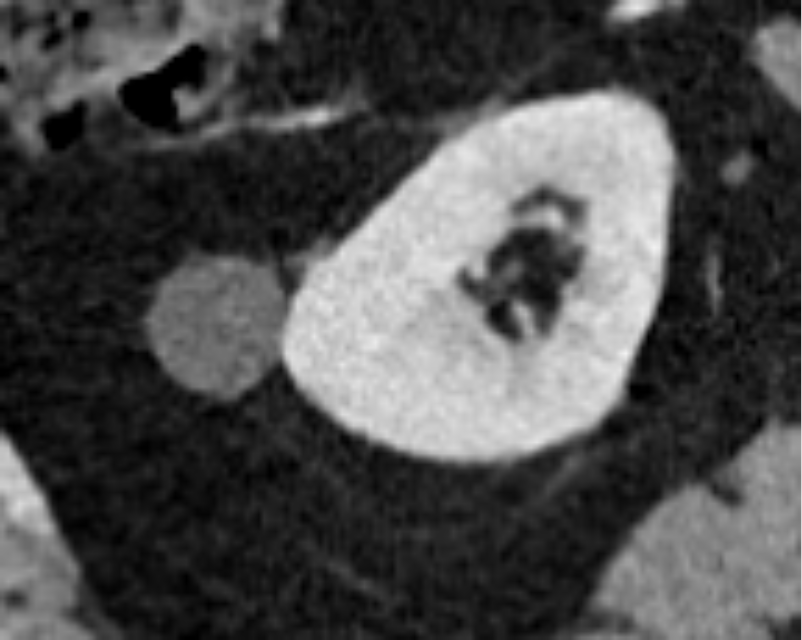}
    \smallskip
    \includegraphics[width=\linewidth,height=2.6cm]{Figures/fig_results/00267_20210602100010_7_N197_Slice_376_Unlabelled.png}
    \smallskip
    \includegraphics[width=\linewidth,height=2.6cm]{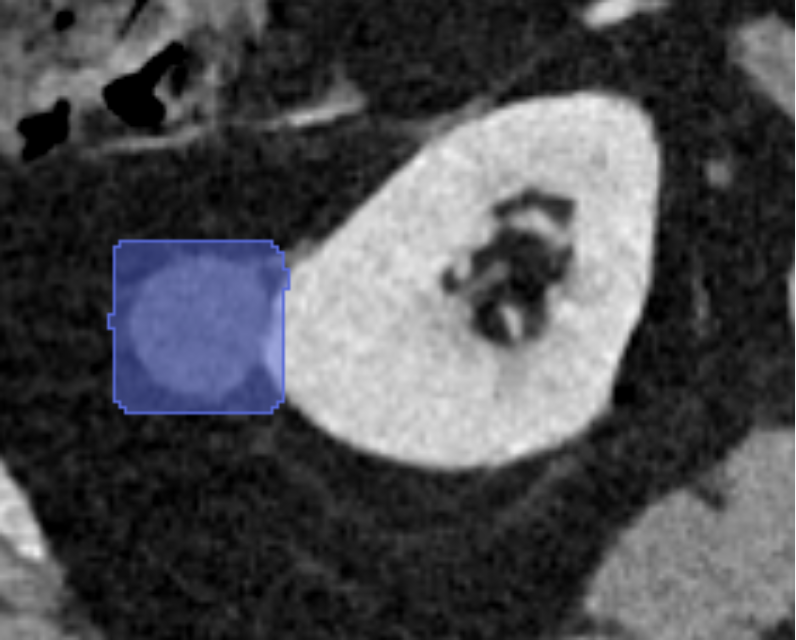}    
    %\caption{}    
\end{subfigure}

\caption{Rows 1, 2, and 3 show cropped CT slices, ground-truth PPGLs (yellow), and detected PPGLs (blue) overlaid, respectively. Columns 1, 2, and 3 show true positives; a small part of an oblong tumor in column 3 was detected by nnUNet. In Column 4, a false positive incorrectly predicted by nnUNet is shown.}

\label{FIGURE LABEL}

\end{figure}

\section{Experiments and Results}
\label{sec:exp_results}

\textbf{Implementation.} nnUNet was trained using 5-fold cross-validation with the different initialization of trainable parameters for a total of 1000 epochs. The loss function used by the model was an equally weighted combination of binary cross-entropy and soft Dice losses. It was optimized using the Stochastic Gradient Descent (SGD) optimizer with an initial learning rate of $10^{-3}$ and a batch size of 1. Each CT volume in the test split was passed to the model from each fold, and predictions from five folds were ensembled together. All experiments were done on a workstation running Ubuntu 22.04 LTS with a NVIDIA Tesla V100 GPU.

\noindent
\textbf{Metrics.} We used precision and recall as the evaluation metrics to quantify the performance of our 3D nnUNet model. True positives (TPs) were those ground truth lesions that overlapped with predicted lesions. False positives (FPs) were those predictions that did not overlap with any ground truth. Finally, false negatives (FNs) were ground truth lesions that were missed by our model.

\noindent
\textbf{Experiments.} Upon qualitative evaluation, we noticed that there were many small predictions. To evaluate the degree of model performance when small lesions were excluded, we designed two experiments. In our first experiment, all the lesions predicted by the nnUNet models were used for metric computation. In the second experiment, a size threshold of 250 voxels on the predictions was set to exclude any predicted small lesions. This size threshold was approximately equal to the 15th percentile of lesion sizes predicted by our 3D nnUNet model. Size thresholds greater than 250 voxels removed true positives at higher rates, resulting in a decrease in recall.

%%%%%%%%%%%%%%%%%%%%%%%%%%%%%%%%%%%%%%%%%%%%%%%%%%%%%%%%%%%%%%%%%%%%%%
%%%%%%%%%%%%%%%%%%%%%%%%%%%%%%%%%%%%%%%%%%%%%%%%%%%%%%%%%%%%%%%%%%%%%%
%Table comparing evaluation metrics for different prediction volume size thresholds
\begin{table*}[!t]
\centering\fontsize{10}{12}\selectfont % to make font size 9 pt
\setlength{\tabcolsep}{7pt} % set small spacing between entries of column (default 6pt)
\setcellgapes{3pt}\makegapedcells % small space between row entries (default 3pt)
\caption{Results of PPGL detection with nnUNet for different prediction volume size thresholds.}
\begin{adjustbox}{max width=\textwidth}
\begin{tabular}{l|c|c|c|c|c|c}
\hline
Size Threshold (in voxels) & \# ground-truth & \# TP & \# FP & \# FN & Precision & Recall \\
\hline 
0 & 153 & 98 & 59 & 55 & 62.4\% & 64.1\% \\
250 & 153 & 98 & 42 & 55 & \textbf{70.0\%} & \textbf{64.1\%} \\
\hline
\end{tabular}
\end{adjustbox}
\label{table_res_overall}

\medskip
\medskip

\caption{Results of PPGL detection with our nnUNet model at the patient level for different predicted size thresholds.}
\begin{adjustbox}{max width=\textwidth}
\begin{tabular}{l|c|c|c|c|c|c|c|c}
\hline
\multirow{2}{*}{Metric} & \multicolumn{4}{c|}{Size Threshold 0} & \multicolumn{4}{c}{Size Threshold 250} \\ \cline{2-9}
                        & Mean & {Std. Dev.} & Median & IQR & Mean & {Std. Dev.} & Median & IQR \\ \hline
Precision               & 62.6\% & {37.0\%} & {66.7\%} & {33.3\% - 100.0\%} & {68.9\%} & {36.0\%} & {84.9\%} & {47.5\% - 100.0\%} \\  
Recall                  & {66.7\%} & {40.2\%} & {100.0\%} & {50.0\% - 100.0\%}  & {66.7\%} & {40.2\%} & {100.0\%} & {50.0\% - 100.0\%} \\ \hline
\end{tabular}
\end{adjustbox}
\label{table_res_patient_level}
\end{table*}

\noindent
\textbf{Results.} Quantitative results of our 3D full resolution nnUNet are presented in Tables \ref{table_res_overall} and \ref{table_res_patient_level}. 
As shown in Table \ref{table_res_overall}, with no prediction volume size threshold applied, our 3D full resolution nnUNet model achieved a precision of 62.4\% and a recall of 64.1\% for PPGL detection. Our recall was greater than 50\%, which suggested that our model can detect different types of PPGLs despite their genetic diversity \cite{nolting2022personalized} in our test dataset. From Table \ref{table_res_patient_level}, on a patient level, the median precision was 66.7\% and recall was 100\%, further supporting our model’s ability to detect different PPGLs. Next, we excluded small predictions with a prediction volume size threshold of 250 voxels, and our model now achieved a higher precision (70.0\% vs. 62.4\%) and the same recall (64.1\%). Importantly, at the patient level, the median precision increased from 66.7\% to 84.9\%, an increase of 18.2\%. 

%%% ------------------ ATTENTION ------------------  
%%% ------------------ ATTENTION ------------------  
%%% THIS IS FOR JOURNAL PAPER
% We also calculated these evaluation metrics for CT scans with SDHx (precision = 68.6\%, recall = 61.5\%), sporadic (precision=66.6\%, recall=62.3\%), kinase (precision = 65.2\%, recall = 62.5\%), and VHL/EPAS1 (precision = 50.0\%, recall = 72.4\%) PPGLs. Our model’s performance also improved for CT scans with SDHX, sporadic, SDHX, and VHL/EPAS1 PPGLs. For CT scans with VHL/EPAS1 PPGLs, precision increased from 50.0\% to 63.6\% while recall remained 72.4\%. For kinase PPGLs, precision increased from 65.2\% to 68.2\% while recalled remained 62.5\%. For sporadic PPGLs, precision increased from 68.6\% to 72.7\% while remained 61.5\%. For SDHX PPGLs, precision increased from 66.7\% to 73.1\% and recall increased from 61.5\% to 62.3\%.

\section{Discussion and Conclusion}

In this work, we proposed to automatically detect PPGLs in CT scans via a proxy segmentation task using the 3D full-resolution nnUNet model. With a prediction volume size threshold of 250 voxels, our model attained a 70.0\% precision and 64.1\% recall. The exclusion signified that the discarded predictions were actually small false positives as opposed to false negatives. Our results suggested that the weak 3D box annotations were sufficient to train a 3D nnUNet model, and the model can potentially achieve clinically acceptable results for PPGL detection in CT scans. The PPGLs in our dataset were of various genetic makeup, such as sporadic, SDHX, Kinase and VHL/EPAS1 \cite{nolting2022personalized}, and they were all obtained at a single institution. We did not delve deeper into the detection results for each genetic cluster and this is a limitation of our work. Another limitation is related to the generation of the ground-truth. A consequence of deriving the weak 3D box-based annotations for PPGLs was the over-estimation (for small tumors) and under-estimation (for large tumors) of the true 3D extent of a tumor. The nnUNet model also inherited these biases post the completion of the training process. We noticed certain instances where the prediction would disappear after 7 slices despite the tumor still persisting for additional slices. We also noticed cases where predictions exhibited box-like appearances that resembled our box-based annotations. Precise delineation of the tumors would circumvent these problems and enable the nnUNet model to be trained effectively to segment the PPGLs for their entire extent. Despite these limitations, to the best of our knowledge, our pilot work is the first to detect PPGLs in CT volumes via a proxy segmentation task.

% Our model attained a precision rate of 62.4\% and a recall rate of 64.1\%, all without considering any threshold on PPGLs' volume size. The model's performance further improved to a precision of 70.0\% while the recall rate was maintained at 64.1\% when a threshold on the PPGL prediction by applying a threshold on PPGLs' volume size, achieving a . Our findings highlight the potential of a deep learning framework to facilitate the automatic detection of PPGLs.  

\appendix    %>>>> this command starts appendixes

\noindent
% \acknowledgments % equivalent to \section*{ACKNOWLEDGMENTS}       
\textbf{Acknowledgements.} This work was supported by the Intramural Research Programs of the NIH Clinical Center and NICHD. The work utilized the computational resources of the NIH HPC Biowulf cluster.

% References
% \clearpage
\bibliography{references} % bibliography data in report.bib
\bibliographystyle{spiebib}

\end{document}